\documentclass{article}
\usepackage{graphicx} % Required for inserting images
\usepackage{amsmath}
\usepackage{amssymb}
\usepackage{authblk}
\usepackage{tcolorbox}
\usepackage{hyperref}
\definecolor{boxcolor}{RGB}{2, 122, 255}

\title{MP-You \\ A Web-based MPI Simulation Tool}
\author[1]{The-Vinh Tran-Luu}
\author[1,2]{Mark-Alexander Henn}
\author[2]{Klaus N. Quelhas}
\author[2]{Solomon I. Woods}
\affil[1]{University of Maryland, College Park, MD, 20742, USA}
\affil[2]{National Institute of Standards and Technology (NIST), Gaithersburg, MD, 20895, USA}
\date{\today}
\begin{document}

\maketitle

\begin{abstract}
Magnetic particle imaging (MPI) is an emerging imaging technique with many applications and a very active field of research. This app provides users with the opportunity to develop some intuition about the inner workings of MPI as it is being researched through NIST's Thermal MagIC project in an interactive and fun way. Users can vary different experimental and post-processing parameters to see how the image quality and particle reconstruction changes for different measurement conditions. 
\end{abstract}

\section{Introduction}
Magnetic particle imaging (MPI) was introduced almost twenty years ago as a way for remotely detecting magnetic nanoparticle (MNP) tracers, with several applications in biomedical imaging and diagnosis, as well as materials research \cite{weizenecker2007simulation,Knopp2012,Gleich2005}. MPI relies on the non-linear magnetization response of MNPs when exposed to time-varying magnetic fields. By generating a moving field-free point (FFP) which saturates all particles but the ones near the no-field region we can scan a specified field of view and measure the magnetization response of the particles in the region of the FFP to a time-varying excitation field.

%establish a connection between the spatial distribution of particles and the measured time-signal. 
%through the generation of a moving field-free point (FFP) which saturates all particles but the ones near the no-field region.
%Special arrangements of magnets or electromagnets allow selectively saturating the particles' magnetization through the generation of a moving field-free point (FFP) over time.
%over time in such a way that the distribution of MNPs can be measured as a function of space, through the generation of a moving field-free point (FFP) which saturates all particles but the ones near the no-field region.   

However, the time signal of the varying magnetic response alone is not enough to deduce the spatial distribution of the MNPs. Obtaining this information requires two additional steps, in which we first relate the time signal to its corresponding spatial position to get an image, and then deblur this image based on the knowledge of the experimental setup and its point spread function (PSF). This process in mathematical terms is also known as solving an inverse problem and amounts to carefully balancing between a good fit to the observed data and a realistic reconstruction.

This web application intends to give the user a tool to see for themselves how different experimental setups can influence the quality and reliability of Magnetic Particle Imaging and can help to illustrate the challenges researchers at NIST face when trying to improve this novel measurement technology\cite{bui2023harmonic,henn2022improving,natorf2024gpu}. It employs a simplified version of NIST's MPI library for Python\footnote{Reference is made to commercial products to adequately specify the experimental procedures involved. Such identification does not imply recommendation or endorsement by the National Institute
of Standards and Technology, nor does it imply that these products are the best for the purpose
specified.}, and runs using Python Flask, HTML, and CSS, as well as matplotlib and other dependencies. The project code is avaliable for viewing and download on \href{https://github.com/xilacxilac/ThermalMagicWebsite}{\textbf{GitHub}}. You can run the project by cloning the repository, installing the dependencies in the \texttt{requirements.txt} text file, and running the command "python app.py" in the project directory. Alternatively, the user can access a web-based version of the app via \href{http://thermalmagic.pythonanywhere.com/}{\textbf{this link}}.

Similar to MPI, the app consists of two major parts: the signal generation, and the reconstruction of the particle distribution from the generated signal.

\subsection{Signal Generation}
The signal measured in an MPI setup can be modeled using:
\begin{equation}\label{Eq:Mod}
    s(t)=-B_s\frac{d}{dt}\int_{\Omega}\rho(x)m \mathcal{M}_{\textsf{eff}}[H_{\textsf{app}}(x,t)]dx,
\end{equation}
with $\rho$ the particle distribution, $m$ is the magnetic moment of a single particle given by the product of the saturation magnetization $M_s$ in [A/m] and the volume of the core of a single particle, that itself depends on the particle's diameter $d$, $B_s$ is the coil sensitivity given in [T/A], and $\mathcal{M}_{\textsf{eff}}$ is a function describing the magnetization response to a time-varying applied field $H_{\textsf{app}}$. 

A commonly applied model for this magnetization behaviour is:
\begin{eqnarray}
    \mathcal{M}_{\textsf{eff}}[H_{\textsf{app}}(x,t)]&=&\mathcal{L}[\kappa \|H_{\textsf{app}}(x,t)\|]\frac{H_{\textsf{app}}(x,t)}{\|H_{\textsf{app}}(x,t)\|},
\end{eqnarray}
with the so-called Langevin function
\begin{eqnarray}    
    \mathcal{L}[\kappa \|H_{\textsf{app}}(x,t)\|]&=&\coth[\kappa \|H_{\textsf{app}}(x,t)\|]-\frac{1}{\kappa \|H_{\textsf{app}}(x,t)\|},
\end{eqnarray}
and $\kappa=\frac{\mu_0 m}{k_B T}$, $k_B$ is the Boltzmann constant and $T$ is the temperature in $K$. MPI utilizes the non-linear magnetization response by creating a field-free point (FFP) that is sensitive to high-frequency changes of the applied field. This is done by combining a gradient field with a gradient $G$ matrix, and a so-called drive field $H_d$, such that the position of the FFP is given by
\begin{equation}\label{Eq:FFP}
    r(t)=G^{-1}H_d(t).
\end{equation}
In the present case, the drive field is defined via
\begin{equation}
    H_d=\begin{bmatrix}
           H_0 \sin(2\pi f_0 t) \\
           H_1 \sin(2\pi f_1 t) 
         \end{bmatrix},
\end{equation}
which leads to a so-called Lissajous trajectory for the FFP.

From Eq. (\ref{Eq:FFP})  it follows that the combination of the gradient field and the drive-field amplitudes determines the maximum range of the trajectory $r$, and hence the dimensions of the field of view (FOV). In the present case, we fix the dimensions of the FOV, such that the drive field amplitudes $H_0$ and $H_1$ become dependent variables. Using the information about the FFP trajectory, we can correlate the time signal from Eq. (\ref{Eq:Mod}) with its spatial position and obtain what sometimes is called the raw image: a blurred version of the original particle distribution, with the degree of blurring determined by the experimental parameters, and ultimately the point spread function of the system. 

\subsection{Image Reconstruction}
The MPI system is a linear-shift invariant system \cite{lu2013linearity}. Hence, the $m$-dimensional signal vector $y$ stemming from an arbitrary distribution of particles $\rho$ can be understood as a weighted sum of signals from $\delta$-samples placed at different positions within the FOV that has been discretized into $n$ voxels. The weights are proportional to the number of particles at that position, and can be determined from a simple matrix equation:
\begin{equation}\label{Eq:MatEq}
    S\rho=y+\epsilon,~\mathrm{with}~S\in\mathbb{R}^{m\times n}, \rho\in\mathbb{R}^n, y\in\mathbb{R}^m.
\end{equation}
The matrix $S$ is called the system matrix, its columns correspond to the measured $m$-dimensional signal from a $\delta$-sample put at the $n$ different positions within the FOV, the term $\epsilon$ denotes possible noise in the measurement. In general $n<m$, which means we are faced with an overdetermined system, so the inverse of the matrix $S$ can not be defined in a straight-forward manner, and instead of solving for $\rho$ directly we need to consider the optimization problem \cite{tarantola2005inverse}:
\begin{equation}\label{Eq:Opt}
    \hat{\rho}=\underset{\rho\in\mathbb{R}^n}{\mathrm{argmin}}\|S\rho-(y+\epsilon)\|.
\end{equation}
Since this system of linear equations is overdetermined, there can be more than one solution that fits the measurement data well. It is however reasonable to expect the solution to this equation to be somewhat regular or smooth, or in other words to have a small $l^2$-norm, leading us to add a so-called regularization term, given by $\lambda_r\|\rho\|$ to our cost function. This leads to the regularized version of our problem:
\begin{equation}\label{Eq:Reg}
    \hat{\rho}=\underset{\rho\in\mathbb{R}^n}{\mathrm{argmin}}\left\{\|S\rho-(y+\epsilon)\|+\lambda_r\|\rho\|\right\}.
\end{equation}
By varying the value of $\lambda_r$ we can balance between the goodness-of-fit to the signal and the desired smoothness of the solution.

\section{User Guide}
Our web-based app employs the models and techniques introduced in the previous section to provide an interactive environment that can help to develop a better understanding of the MPI framework. This process again consists of two separate steps: the signal generation and the reconstruction.

\subsection{Signal Generation in App}
\begin{figure}[h!]
    \centering
    \includegraphics[width=0.99\linewidth]{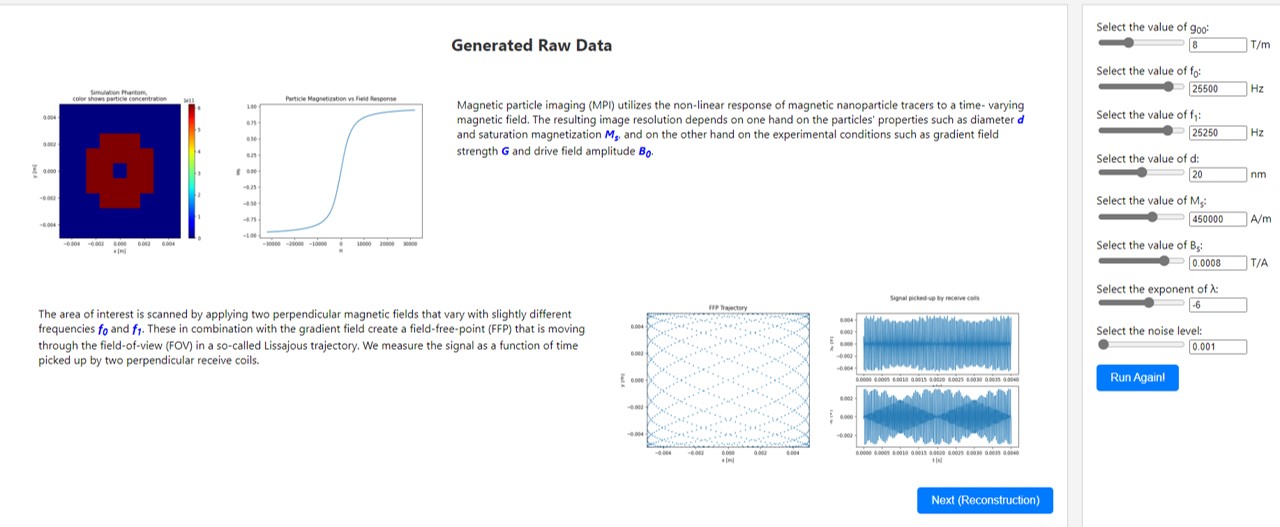}
    \caption{Signal generation screen as seen in app.}
    \label{fig:screen1}
\end{figure}
\begin{figure}[h!]
    \centering
    \includegraphics[width=0.85\linewidth]{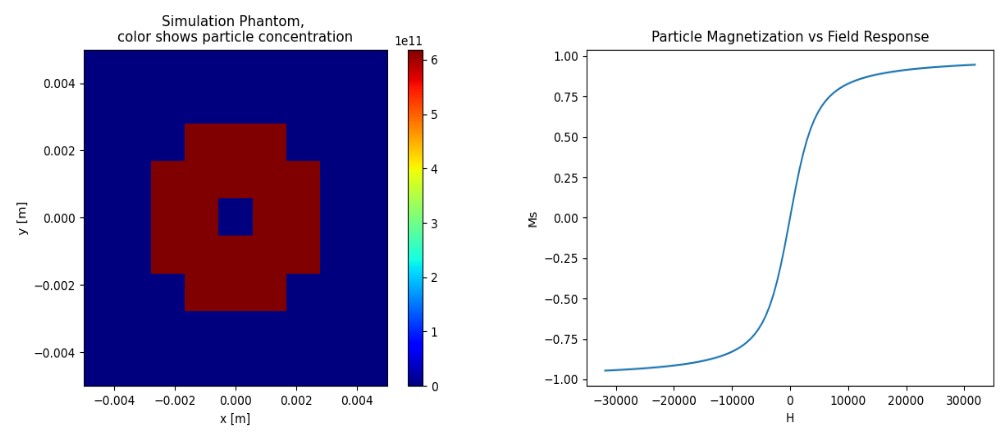}
    \caption{Ground truth particle distribution within FOV (left) and M vs. H magnetization curve (right).}
    \label{fig:plot1}
\end{figure}
\begin{figure}[h!]
    \centering
    \includegraphics[width=0.85\linewidth]{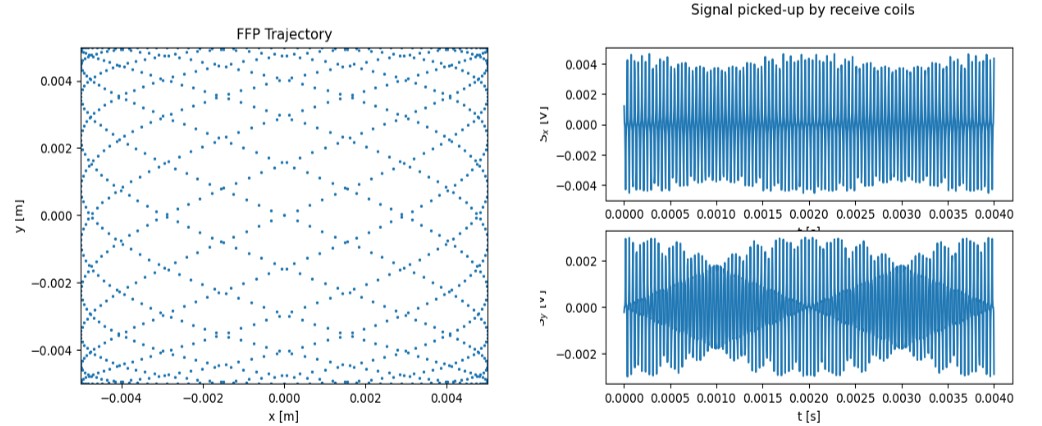}
    \caption{FFP trajectory (left) and raw signals picked up by receive coils (right).}
    \label{fig:plot2}
\end{figure}

The first interactive screen is the simulation screen, in which the raw time signal data can be generated. Once the user hits the \fcolorbox{white}{boxcolor}{\textsf{\textcolor{white}{Run Code!}}} button the app starts calculating the necessary data and puts out four plots. These consist of the ground truth particle distribution and the magnetization curve, see Fig. (\ref{fig:plot1}), and the FFP trajectory and the raw signal, see Fig. (\ref{fig:plot2}). Noise is added to the generated data, via the term $\epsilon$ in Eq. (\ref{Eq:MatEq}). It is drawn from a normal distribution with zero mean and a standard deviation that is proportional to the maximum of the generated signal, hence $\epsilon\sim\mathcal{N}(0,\sigma^2)$, with $\sigma=\kappa\cdot\max(y)$, such that a noise level of $0.01$ corresponds to a standard deviation of 1\% of the maximum of the measured signal.

\begin{table}[ht]
	\centering
	\begin{tabular}{c | c | c}
 \textbf{Parameter} & \textbf{Symbol} & \textbf{Default value} \\
 \hline
		Gradient field in $x$-direction & $g_{00}$ & 8 T/m\\
        Drive coil frequency in $x$-direction & $f_0$ & 25.5 kHz \\
        Drive coil frequency in $y$-direction & $f_1$ & 25.25 kHz \\
        Particle diameter & $d$ & 20 nm \\
        Saturation magnetization & $M_s$ & $4.5\cdot 10^{5}$ A/m\\
        Coil sensitivity & $B_s$ & $8\cdot 10^{-4}$ T/A\\
        Regularization parameter exponent & $\lambda$ & -6 \\
        Noise level & $\kappa$ & $10^{-3}$        
	\end{tabular}
\caption{User adjustable parameters.}\label{tab:Para}	
\end{table}

Note, that the ground truth particle distribution is not affected by varying the experimental parameters, as are the $x$ and $y$ dimensions of the FOV. The M vs. H curve changes with $M_s$, $d$, and $g_{00}$ (see Table \ref{tab:Para}) such that an increase in $M_s$ or $d$ leads to a steeper shape, and a change in only $g_{00}$ only changes the range of $H$ over which we plot the M vs. H curve. It is important to note that the user only provides a single value $g_{00}$ for the gradient matrix G, from we which construct the gradient matrix as:
\[
G=\begin{bmatrix}
g_{00} & 0 \\ 0 & -\frac{1}{2}g_{00}
\end{bmatrix}
\]

In our setting, the FFP trajectory is only a function of $f_0$ and $f_1$, since we fix the FOV and change the drive field amplitudes accordingly, while the raw signal depends on all mentioned parameters, and the noise level. Note, that changing the value of the coil sensitivity $B_s$ only scales the generated signal.

By clicking on the \fcolorbox{white}{boxcolor}{\textsf{\textcolor{white}{Next (Reconstruction)}}} button, the user advances to the next screen which provides the reconstructed particle distribution along with some additional information.

\subsection{Image Reconstruction in App}
\begin{figure}[h!]
    \centering
    \includegraphics[width=0.99\linewidth]{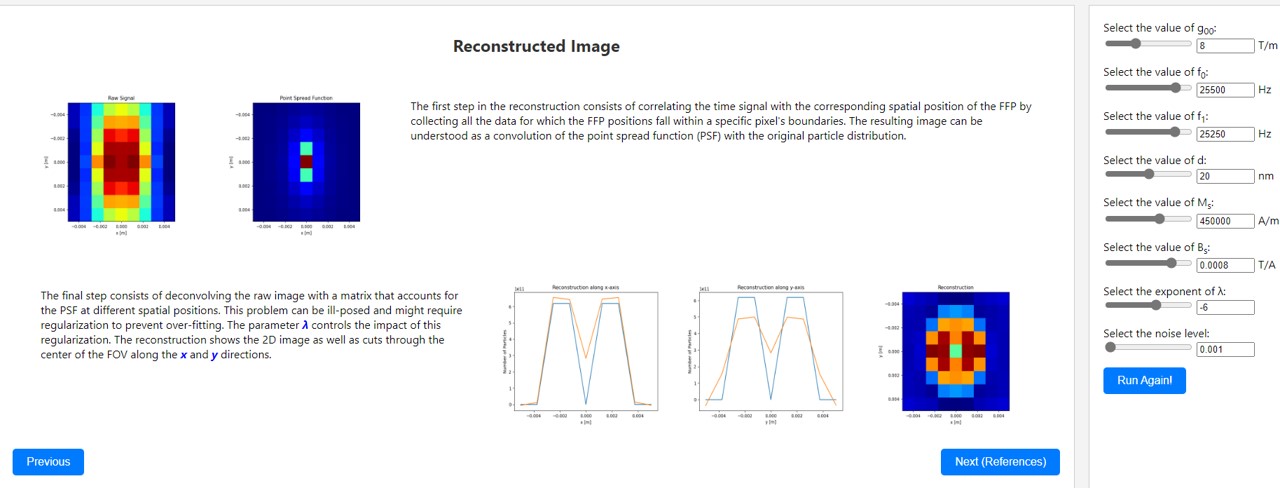}
    \caption{Image reconstruction screen as seen in app.}
    \label{fig:screen2}
\end{figure}
\begin{figure}[h!]
    \centering
    \includegraphics[width=0.85\linewidth]{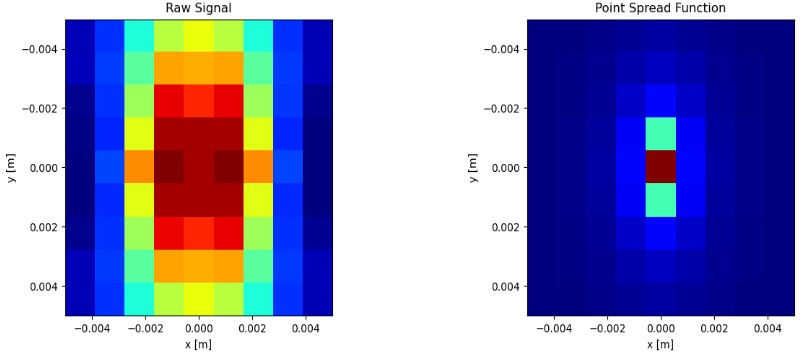}
    \caption{Raw image data (left) and PSF (right).}
    \label{fig:plot3}
\end{figure}
\begin{figure}[h!]
    \centering
    \includegraphics[width=0.85\linewidth]{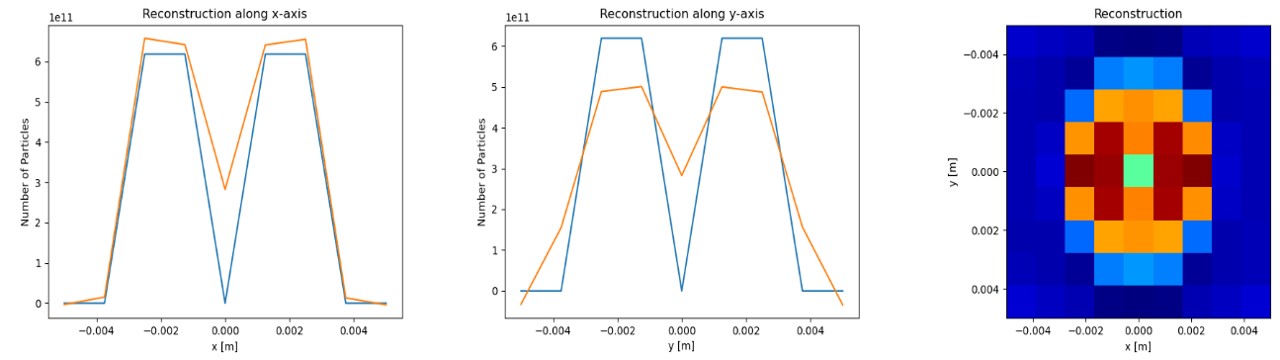}
    \caption{Reconstructed particle distribution, profiles along the $x$- direction (left), $y$- direction (center), and 2D representation (right).}
    \label{fig:plot4}
\end{figure}

Using the data generated in the previous screen, we determine the particle distribution from the raw signal data, and are presented with five plots. The first two in Fig. (\ref{fig:plot3}) show the time signal related to its spatial position, also called the raw image, and the point spread function (PSF) which gives us an idea about the blurring that stems from the selected experimental parameters. In general a larger value for $g_{00}$ leads to less dispersed PSF, and hence a better resolution in the raw image. The second set of plots, Fig. (\ref{fig:plot4}), show the reconstructed particle distribution in terms of profiles along the two FOV axes, and a 2D representation of the reconstruction.

When you hit the \fcolorbox{white}{boxcolor}{\textsf{\textcolor{white}{Run Again!}}} button the reconstruction is rerun. This is helpful to investigate how the regularization parameter $\lambda$ or various experimental parameters change the reconstruction. Note, that the effective regularization parameter $\lambda_r$ in Eq. (\ref{Eq:Reg}) is calculated as $\lambda_r=10^{\lambda}$. 

%\section{Outlook ?}
%Talk about the potential to change the hard-coded variables, such as the ground truth phantom, the dimensions of the FOV, and others by changing the MPIlib python library. However, the user should be aware that manipulating these variables can have severe consequences on the code in general and should be done carefully.

%\section{Changes to Code}
%\begin{itemize}
%\item Use consistent color maps for the 2D plots, e.g., "jet" DONE
%\item Change the title for the 3rd screen to "Reconstruction" DONE (need title for first slide)
%\item Add a default screen IP (not sure what to add for the default screen)
%\item Maybe have a glossary or an active cursor that tells users what the different varibles stand for or an help button DONE
%\item Add units to the variables DONE
%\item Consistent names for frequencies, $f_0$ and $f_1$ or $f_1$ and $f_2$ DONE
%\item Make the FFP and phantom plot square NEED CLARIFICATION
%\end{itemize}

\bibliographystyle{abbrv}
\bibliography{ref}

\end{document}